\documentclass{PoS}

\usepackage{heppennames2}

\usepackage{lineno}

\graphicspath{{./plots/}}

\title{Exclusive Top Threshold Matching at Lepton Colliders}

\ShortTitle{Exclusive Top Threshold Matching at Lepton Colliders}

\author{
  \speaker{J\"urgen Reuter} \\
  DESY, Hamburg, Germany\\
  E-mail: \email{juergen.reuter@desy.de}
}

\author{Bijan Chokouf\'e Nejad \\
DESY, Hamburg, Germany \\
E-mail: \email{bijan.chokoufe@desy.de}
}

\author{Andre Hoang \\
University of Vienna, Vienna, Austria \\
E-mail: \email{andre.hoang@univie.ac.at}
}

\author{Wolfgang Kilian \\
University of Siegen, Germany \\
E-mail: \email{kilian@physik.uni-siegen.de}
}

\author{Maximilian Stahlhofen \\
Johannes-Gutenberg-University, Mainz, Germany \\
E-mail: \email{mastahlh@uni-mainz.de}
}

\author{Thomas Teubner \\
University of Liverpool, United Kingdom \\
E-mail: \email{thomas.teubner@liverpool.ac.uk}
}

\author{Christian Weiss \\
DESY, Hamburg, Germany \\
E-mail: \email{christian.weiss@desy.de}
}

\abstract{
  The threshold scan at future lepton colliders is the most precise
  known method to determine the top quark mass (well below 100 MeV), a
  fundamental parameter of the Standard Model that co-determines the
  stability properties of the electroweak vacuum. We present a new method to
  match the continuum next-to-leading order QCD corrections with the
  next-to-leading logarithmic resummation of the Coulomb singularities of the
  quasi-toponium bound state at threshold where fixed-order perturbation
  theory is invalid. This matching is performed at the level of the fully
  exclusive WbWb final state. It allows to study all kinds of differential
  distributions at or close to threshold. The top mass dependence of these
  distributions opens up new possibilities for the top mass determination
  that might be competitive with the inclusive threshold scan.  
}

\FullConference{39th International Conference on High Energy Physics\\
		4-11 July 2018\\
		Seoul, Korea}

\begin{document}

\section{Introduction}

Future high-energy lepton colliders will measure the top quark
properties like its mass, width and couplings with an unprecedented
accuracy and use the top quark as a means to search for new physics
beyond the Standard Model. Here, we discuss the matching of
fixed-order QCD next-to-leading order (NLO-QCD calculations for
exclusive $e^+e^- \to W^+ b W^- \bar{b}$ final states in the
continuum, based on~\cite{Nejad:2016bci}, with a resummed calculation
in the threshold region, where fixed-order perturbation theory in the
strong coupling $\alpha_s$ is not a good approximation anymore, but
the top velocity $v$ is an additional expansion parameter and
Coulomb-singular terms $\sim (\alpha_s/v)^n$ and (ideally also) large
logarithms $\sim(\alpha_s \log v)^n$ have to be resummed. In
Sec.~\ref{sec:matching}, after reviewing our calculational framework
and the details of the continuum calculation for completeness, we
discuss, based on~\cite{Bach:2017ggt}, a previously known
non-relativistic QCD (NRQCD) effective field theory setup to compute a
form factor accounting for the resummation of the threshold-singular
terms at NLL accuracy, implemented it in the fixed-order calculation
and matched the result to the QCD-NLO cross section in the transition
region between threshold and continuum. We thus obtained a
fully-differential cross section, which gives reliable predictions for
all center-of-mass energies. Depending on how inclusive the process
is, we achieve LL + QCD-NLO (for very exclusive processes) or NLL +
QCD-NLO precision (for inclusive processes) in the threshold
region. Finally, we conclude in Sec.~\ref{sec:conclusions}. 


\section{QCD-NLO (fixed-order) \& Threshold Matching}
\label{sec:matching}

In the continuum, i.e. away from the threshold, QCD corrections are
properly described by fixed-order relativistic QCD-NLO perturbation
theory for the off-shell top pair production. For that purpose, we
study either the process $e^+e^- \to W^+ b W^- \bar{b}$ or $e^+e^- \to
\ell^+ e^- \bar{\nu}_e \mu^+\nu_\mu b \bar{b}$ including leptonic $W$
decays. Within the full four- or six-particle final state, there
are double-resonant diagrams included (involving a top and an anti-top
propagator), single-resonant diagrams and non-resonant irreducible
background processes. To calculate total and fully differential
QCD-NLO corrections for the top production processes, we take the
\texttt{WHIZARD} framework for (QCD-)NLO
processes. \texttt{WHIZARD}~\cite{Kilian:2007gr} is a 
multi-purpose event generator with its own matrix-element
generator for tree-level amplitudes,
\texttt{O'Mega}~\cite{Moretti:2001zz,Nejad:2014sqa} with support
for a plethora of models like
e.g. supersymmetry~\cite{Ohl:2002jp}. Users can use external models by
the interface to
\texttt{FeynRules}~\cite{Christensen:2010wz}. \texttt{WHIZARD} uses
the color-flow formalism~\cite{Kilian:2012pz}, and it comes with its
own parton shower implementation~\cite{Kilian:2011ka}. QCD-NLO
applications within \texttt{WHIZARD} started with a hard-coded
implementation for the production of $b$ jets at
LHC~\cite{Binoth:2009rv,Greiner:2011mp}, while matching 
between resummed terms and fixed-order calculations have been tackled
by combining fixed-order electroweak corrections to chargino
production at the ILC with an all-order QED initial-state structure
function~\cite{Kilian:2006cj,Robens:2008sa}. \texttt{WHIZARD} is also
able to do automatic POWHEG matching for $e^+e^-$ 
processes~\cite{Reuter:2016qbi}.

\texttt{WHIZARD} uses FKS subtraction~\cite{Frixione:1995ms} and
generates the automatically generates the phase space for all singular
emission regions. Virtual matrix elements, color-correlated and
spin-correlated matrix elements for the collinear and soft splittings
are taken from the one-loop provider (OLP) program
\texttt{OpenLoops}~\cite{Openloops}. The complex mass scheme is used,
leading to a complex weak mixing angles. The input values are as follows: $m_W =
80.385$ GeV, $m_Z = 91.1876$ GeV, $m_t = 173.2$ GeV, $m_H = 125$
GeV. We use massive $b$-quarks of mass $m_b = 4.2$ GeV. Widths need to
be calculated at the same order and in the same scheme than the
scattering process in order to guarantee properly normalized branching
ratios: $\Gamma_Z^{\text{LO}} = 2.4409$ GeV,  $\Gamma_Z^{\text{NLO}} =
2.5060$ GeV, $\Gamma_W^{\text{LO}} = 2.0454$ GeV,
$\Gamma_W^{\text{NLO}} = 2.0978$ GeV,  $\Gamma_{t\to Wb}^{\text{LO}} =
1.4986$ GeV, $\Gamma_{t\to Wb}^{\text{LO}} = 1.3681$ GeV. As the
matrix elements for the full off-shell processes contain narrow
resonances, particularly the $H\to bb$ resonance, we use a
resonance-aware version of the FKS subtraction formalism to make sure
that cancellations between real emissions and subtraction terms do
cancel though the real emission could shift the kinematics on or off
the resonance compared to Born kinematics. This resonance-aware
treatment is automatically done in \texttt{WHIZARD}. As we are using
massive $b$-quarks, no cuts are necessary for the  process $e^+e^- \to
W^+W^- b \bar{b}$. The integrations for the full QCD-NLO are
very stable. We did two independent own integrations with the serial
and the non-blocking MPI-parallelizable version~\cite{MPI} of
\texttt{VAMP}~\cite{Ohl:1998jn} inside \texttt{WHIZARD}. 

For the QCD-NLO corrections, we take the top mass as renormalization
scale. The scale variations for the process $e^+e^- \to W^+ b W^-
\bar{b}$ is very small, at the level of two per cent. After one has
replaced the top width in the matrix elements by a running top width
$\Gamma_t(\mu_R)$ , the scale variations for the on-shell process
$e^+e^- \to t\bar{t}$ behave the same way as for the off-shell
process. 
The \texttt{WHIZARD} infrastructure immediately enables QCD-NLO
calculations/simulations for polarized beams, to include QED
initial-state photon radiation as well as collider-specific 
beamspectra.

A kinematic fit to the shape of the rising of the cross section at the
top threshold is believed to be the most precise method to measure the
top quark mass with an ultimate precision of 30-80 MeV. For this the
systematic uncertainties of the experimental measurement -- especially
the details of the beam spectrum -- as well as the theoretical
uncertainties have to be well under control. As shown above, close to
the kinematical threshold for the on-shell production of a $t\bar{t}$ 
\begin{figure}
  \begin{center}
    \includegraphics[width=.44\textwidth]
                    {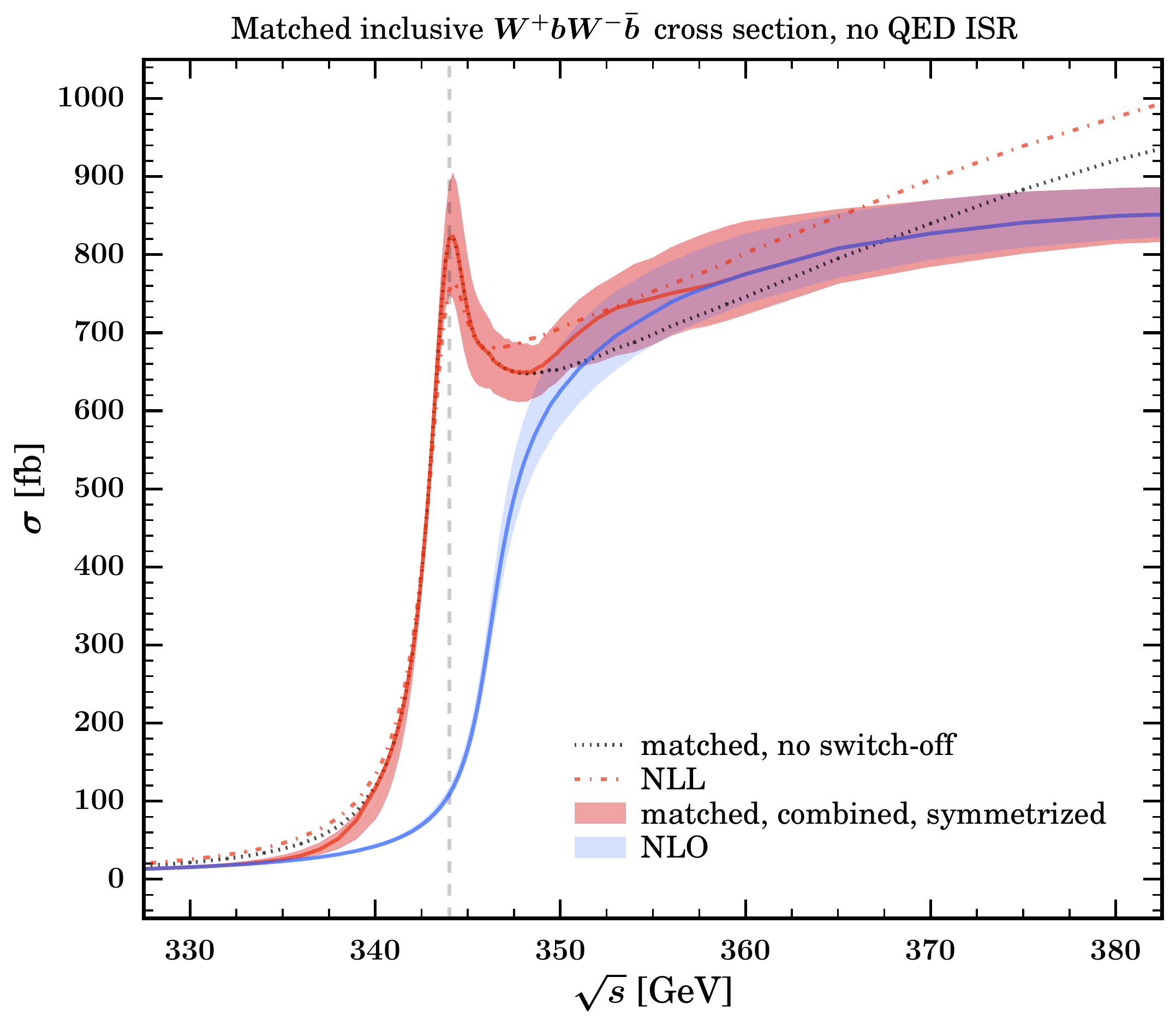}
                    \quad
    \includegraphics[width=.44\textwidth]
                    {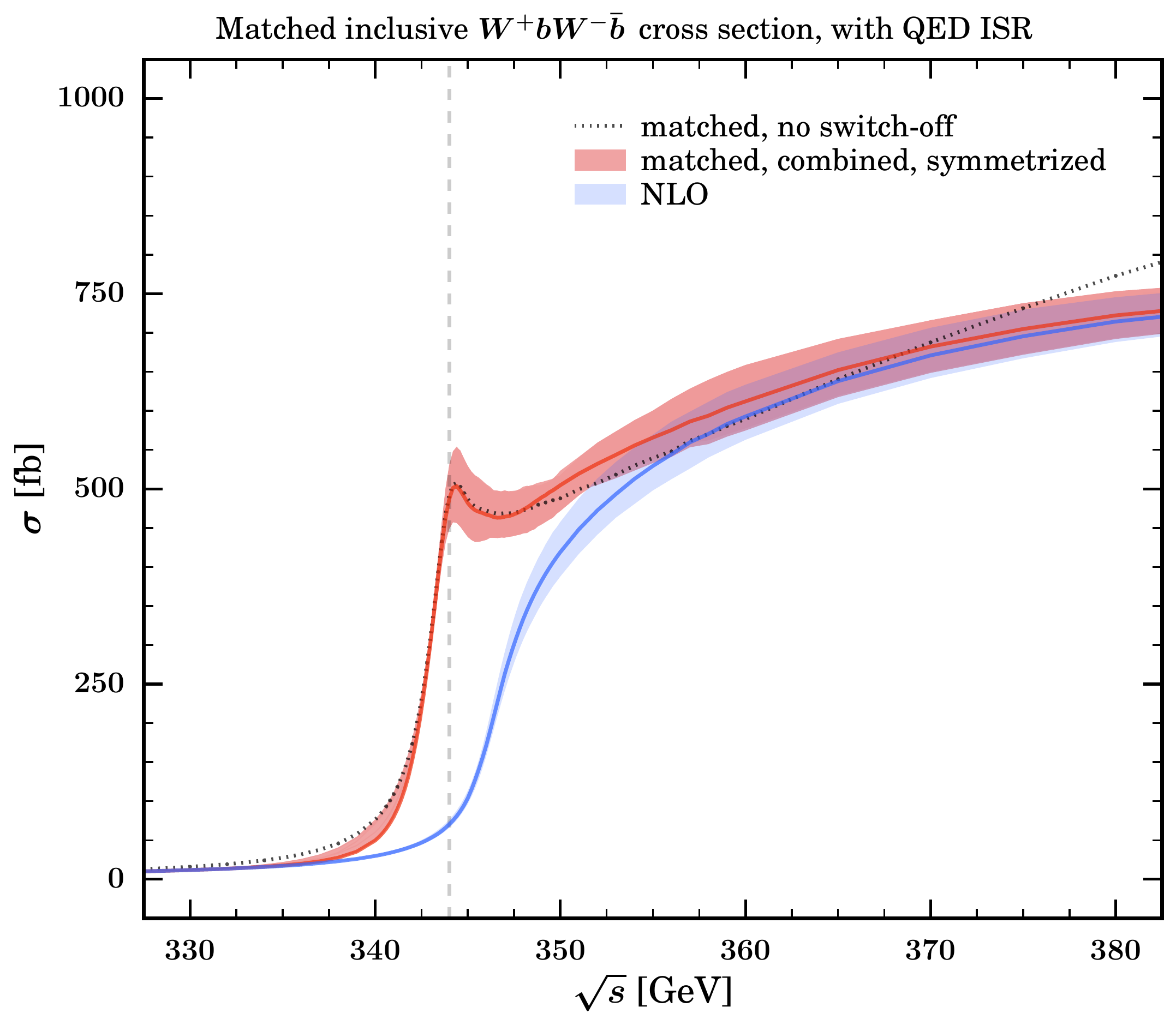}    
  \end{center}
  \caption{Matched NRQCD-NLL + QCD-NLO 
    calculation without (left) and with (right) QED ISR. The dashed
    vertical line is the value of twice $M^{1\text{S}}$. Blue is the
    fixed QCD-NLO calculation, red is the fully matched
    calculation. The matched calculation has a full envelope over
    (symmetrized) scale uncertainties as well as variations over
    switch-off functions. }
  \label{fig:threshold_full}  
\end{figure}
pair, fixed-order perturbation theory is not a good
approximation. Very close to threshold, the effective field theory of
(v/p)NRQCD separates the hard scale $m_t$, the soft scale given by the
top momentum of the non-relativistic top quark with velocity $v$, $m_t
v$ and the ultrasoft scale, given by the kinetic energy of the top
quark, $m_t v^2$ and allows to resum large logarithms of $v$
with $\alpha_s \sim v \sim 0.1$ close to
threshold. "Fixed-order" calculations resumming only Coulomb
singularities, but no velocity logarithms, for the totally inclusive 
$t\bar{t}$ production have been carried out in NRQCD to
NNNLO~\cite{Beneke:2015kwa}. The large velocity logarithms have been
resummed to next-to-next-leading logarithmic
(NNLL)~\cite{Hoang:2013uda} order
(cf. also~\cite{Hoang:2001mm,Pineda:2006ri} for predictions not
containing the full set of NNLL ultrasoft logarithms). These NRQCD 
calculations, based on the optical theorem, hold only for the
total inclusive cross section in a narrow window around the
$t\bar{t}$ threshold. Here, we combine and match the NLL
NRQCD-resummed process close to the top threshold with 
the fixed-order (relativistic) QCD-NLO process in the continuum. By a
carefully performed matching procedure, our approach smoothly
interpolates between threshold region and continuum, and
allows to study all kinds of differential distributions. 

The matching is embedded into the \texttt{WHIZARD}-\texttt{OpenLoops}
QCD-NLO fixed-order framework discussed above. The NLL resummed NRQCD
contributions are included in terms of (S-/P-wave) form factors to the
(vector/axial vector) $\gamma/Z-t-\bar{t}$ vertex. These form factors
are obtained from the numerical solution of Schr\"odinger-type
equations for the NLL Green functions computed by the 
\texttt{Toppik}~\cite{Jezabek:1992np,Harlander:1994ac,Hoang:1999zc}
code, which is included in \texttt{WHIZARD},  for technical details
cf.~\cite{Bach:2017ggt}. In order to avoid double-counting between 
the fixed-order QCD-NLO part and the resummed NLL-NRQCD part, one has
to expand the form factors to first order in $\alpha_s$ and subtract
those pieces. As the NRQCD resummed calculations are only available
for the top-vector and axial-vector currents, this removal of
double-counting has to be done in a factorized approach within a
double-pole approximation. In order to maintain gauge-invariance of
the factorized amplitudes, an on-shell projection of the exclusive
final states to the top mass shell is performed, for details
cf.~\cite{Bach:2017ggt}. The implementation inside \texttt{WHIZARD}
has been 
\begin{figure}
  \begin{center}
    \includegraphics[width=.40\textwidth]
                    {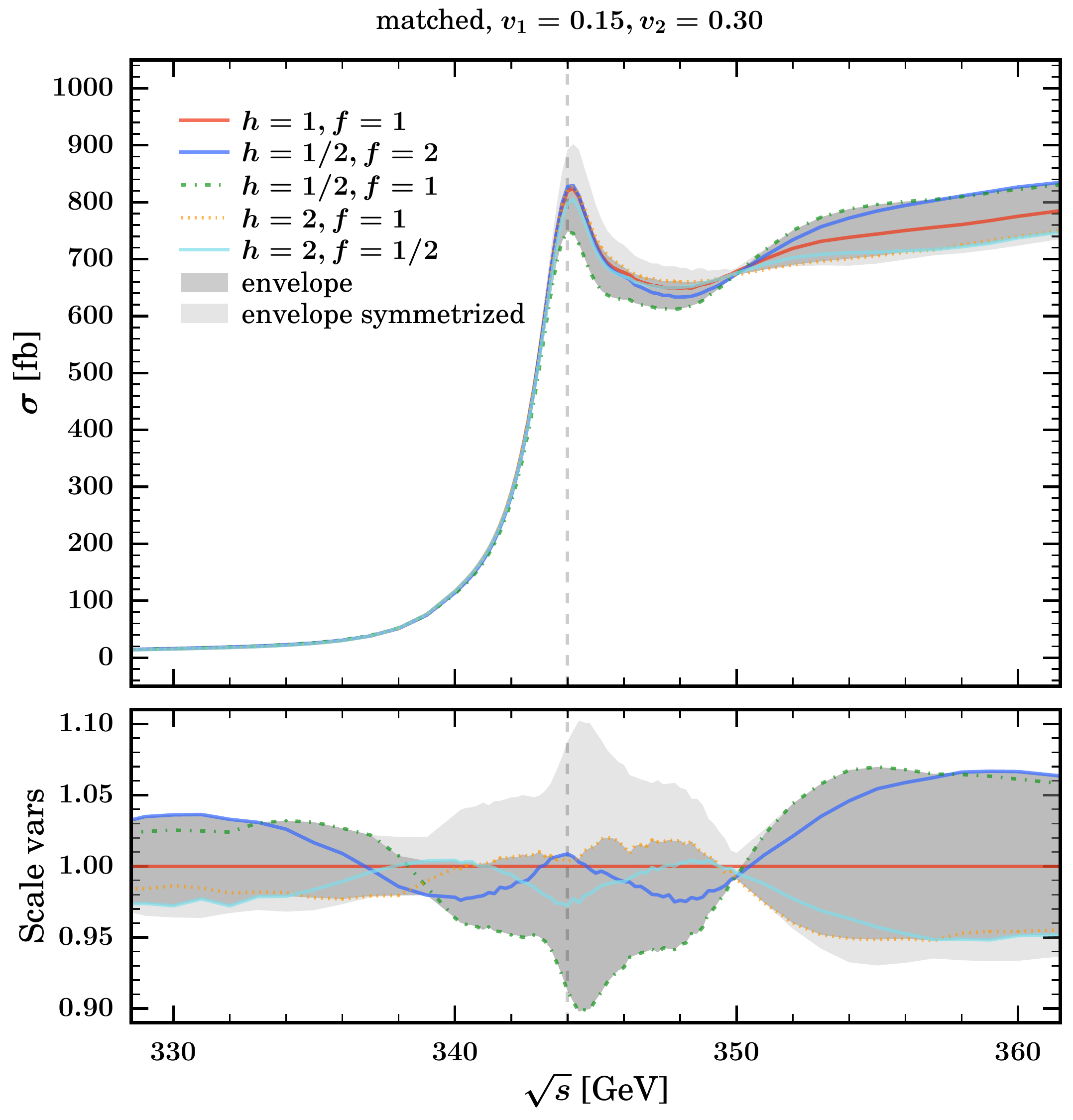}
                    \quad
    \includegraphics[width=.44\textwidth]
                    {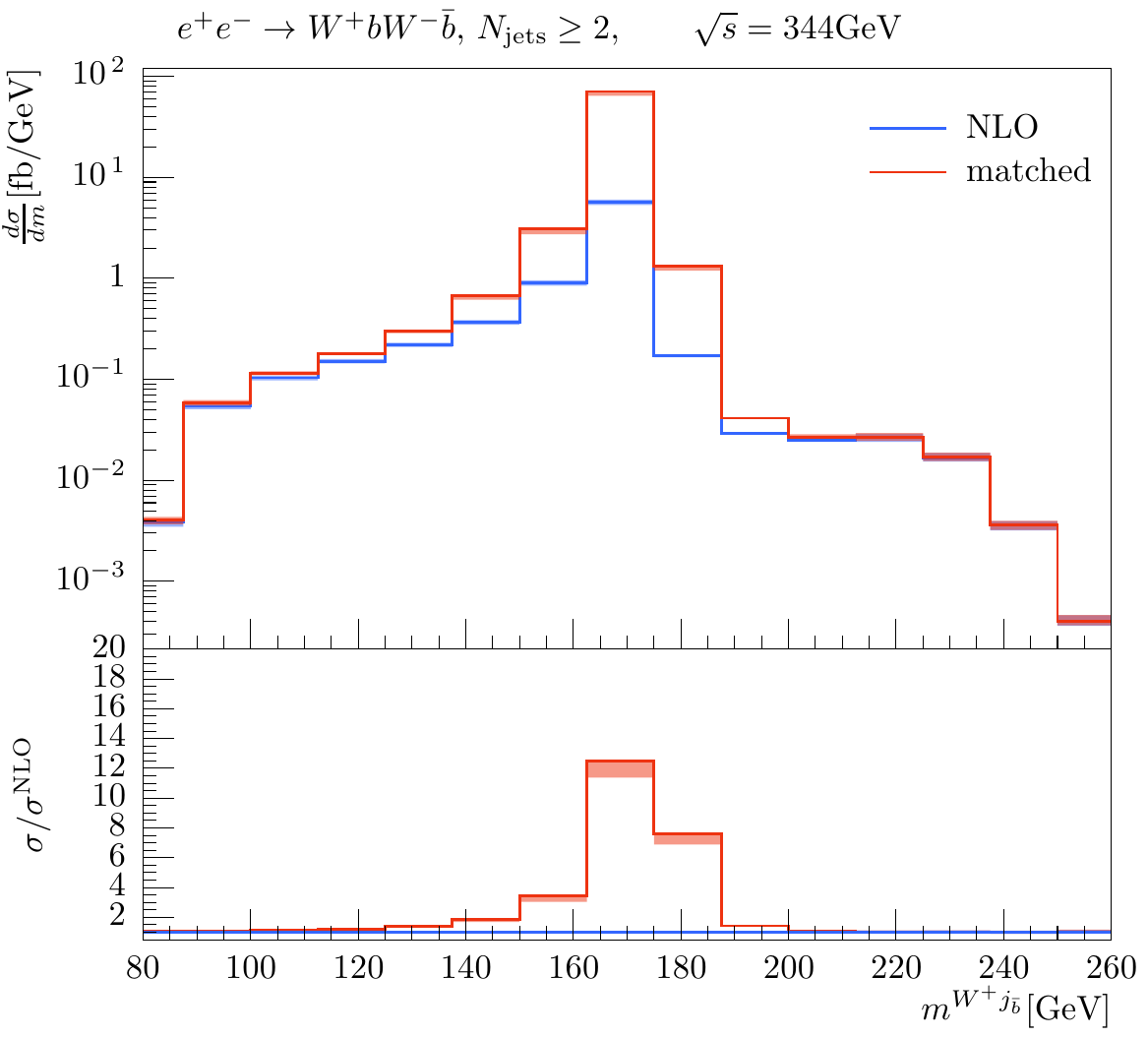}    
  \end{center}
  \caption{Left panel: Matched NRQCD-NLL + QCD-NLO total cross section
    as on the left of Fig.~\ref{fig:threshold_full}, but for a single
    choice of switch off-function. $h$ and $f$ are renormalization
    scale parameters as defined in
    \cite{Bach:2017ggt,Hoang:2013uda}. The grey bands display the
    corresponding scale 
    variations with and without symmetrization. Right panel: $Wb$
    invariant mass distribution at threshold ($\sqrt{s}=344$ GeV) as
    obtained with \texttt{WHIZARD}. The red line represents the full
    NRQCD-NLL + QCD-NLO matched, and the blue line the pure QCD-NLO
    result. The associated bands are generated by the same scale
    variations as in the left panel, here without symmetrization.}
  \label{fig:match_diff}
\end{figure}
validated with analytical calculations for different invariant mass
cuts on the reconstructed top quarks from Ref.~\cite{Hoang:2010gu}.

For larger top velocity ($v \gtrsim 0.4$) only the relativistic
QCD-NLO result is valid. We define a switch-off function that smoothly
interpolates between the two regions. The possibility to vary this
arbitrary function and its parameters adds another theory uncertainty
to the different scale variations~\cite{Bach:2017ggt}. The
results of our matching procedure are displayed in
Fig.~\ref{fig:threshold_full}. These plots 
show the total inclusive cross section for the process $e^+e^- \to W^+ b W^-
\bar{b}$, in the left panel without and in the right panel with QED
initial-state radiation (ISR). The dashed vertical line gives the
value for $2 M^{1\text{S}}$. The 1S mass $M^{1\text{S}}$ is defined as
half of the perturbative mass of a would-be 1S toponium state and
represents a renormalon free short-distance mass, which we treat as an
input parameter in \texttt{WHIZARD}. The blue
line shows the QCD-NLO cross section including scale variations in the
blue shaded areas. The red curve shows the NRQCD-NLL + QCD-NLO
result, while the shaded band contains all (symmetrized) scale
variations of the hard, soft and ultrasoft
factorization/renormalization scales according
to~\cite{Hoang:2013uda} as well as variations of the switch-off
function to a reasonable extent~\cite{Bach:2017ggt}. The dotted black
line shows the matched results without applying a switch-off function
to the factorized NRQCD terms which deviates above threshold from
the relativistic QCD-NLO result. In Fig.~\ref{fig:match_diff}, left
panel, we see the matched result in the threshold region for a
single choice  of switch-off parameters, but scale variations over the
full two-dimensional renormalization parameter range defined
in~\cite{Hoang:2013uda}. This shows that the 
scale variation bands for the resummed NLL result in the threshold
region are highly asymmetric with respect to the central value which
motivates to apply a symmetrization of the error bands around the
central value. This symmetrization is also shown in
Fig.~\ref{fig:threshold_full}. In the right panel of
Fig.~\ref{fig:match_diff} we show as an example for a differential
distribution the invariant mass of the $W-b$ jet system. Blue is the
fixed-order QCD-NLO distribution, while red is the fully matched
distribution including scale variations, here un-symmetrized. The
ratio plot in the bottom does not show a K factor, but the ratio of
the matched result to the QCD-NLO fixed order result. It shows an
enhancement in the top mass peak due to threshold resummation by a
factor of 10-12.

 
\section{Conclusions}
\label{sec:conclusions}

In order to be able to study experimental event selections as well as
differential distributions, we presented a matched threshold
calculation that smoothly interpolates the threshold region
described by non-relativistic QCD to the relativistic QCD-NLO
calculation. It constitutes the highest precision available at the
level of the completely exclusive final state. Any of the presented
differential distributions depending on the top mass may serve as a
different means to determine the top mass. We were not accomplishing
this task here, but rather showed a framework as a proof-of-principle
of the matching procedure between threshold and continuum. For the
proper matching to the continuum the fixed-order QCD-NLO calculations
for top-quark pair production including top and (leptonic) $W$ decays
have been done. All of this has been done in the  QCD-NLO framework of
the \texttt{WHIZARD} event generator which allows to include all
important physics of a lepton collider like polarization, QED ISR
radiation and non-trivial beam spectra.


\begin{thebibliography}{99}

\bibitem{Nejad:2016bci} 
  B.~Chokouf\'e Nejad, W.~Kilian, J.~M.~Lindert, S.~Pozzorini, J.~Reuter
  and C.~Weiss, 
  JHEP {\bf 1612}, 075 (2016)
  [arXiv:1609.03390 [hep-ph]].

\bibitem{Bach:2017ggt} 
  F.~Bach, B.~C.~Nejad, A.~Hoang, W.~Kilian, J.~Reuter, M.~Stahlhofen, T.~Teubner and C.~Weiss,
  JHEP {\bf 1803}, 184 (2018)
  [arXiv:1712.02220 [hep-ph]].

\bibitem{Kilian:2007gr} 
  W.~Kilian, T.~Ohl and J.~Reuter,
  Eur.\ Phys.\ J.\ C {\bf 71}, 1742 (2011)
  [arXiv:0708.4233 [hep-ph]].
  
\bibitem{Moretti:2001zz} 
  M.~Moretti, T.~Ohl and J.~Reuter,
  hep-ph/0102195.

\bibitem{Nejad:2014sqa} 
  B.~Chokoufe Nejad, T.~Ohl and J.~Reuter,
  Comput.\ Phys.\ Commun.\  {\bf 196}, 58 (2015)
  [arXiv:1411.3834 [physics.comp-ph]].
  
\bibitem{Ohl:2002jp} 
  T.~Ohl and J.~Reuter,
  Eur.\ Phys.\ J.\ C {\bf 30}, 525 (2003)
  [hep-th/0212224].

\bibitem{Christensen:2010wz} 
  N.~D.~Christensen, C.~Duhr, B.~Fuks, J.~Reuter and C.~Speckner,
  Eur.\ Phys.\ J.\ C {\bf 72}, 1990 (2012)
  [arXiv:1010.3251 [hep-ph]].

\bibitem{Kilian:2012pz} 
  W.~Kilian, T.~Ohl, J.~Reuter and C.~Speckner,
  JHEP {\bf 1210}, 022 (2012)
  [arXiv:1206.3700 [hep-ph]].

\bibitem{Kilian:2011ka} 
  W.~Kilian, J.~Reuter, S.~Schmidt and D.~Wiesler,
  JHEP {\bf 1204}, 013 (2012)
  [arXiv:1112.1039 [hep-ph]].

\bibitem{Binoth:2009rv} 
  T.~Binoth, N.~Greiner, A.~Guffanti, J.~Reuter, J.-P.~Guillet and T.~Reiter,
  Phys.\ Lett.\ B {\bf 685}, 293 (2010)
  [arXiv:0910.4379 [hep-ph]].
  
\bibitem{Greiner:2011mp} 
  N.~Greiner, A.~Guffanti, T.~Reiter and J.~Reuter,
  Phys.\ Rev.\ Lett.\  {\bf 107}, 102002 (2011)
  [arXiv:1105.3624 [hep-ph]].

\bibitem{Kilian:2006cj} 
  W.~Kilian, J.~Reuter and T.~Robens,
  Eur.\ Phys.\ J.\ C {\bf 48}, 389 (2006)
  [hep-ph/0607127].
  
\bibitem{Robens:2008sa} 
  T.~Robens, J.~Kalinowski, K.~Rolbiecki, W.~Kilian and J.~Reuter,
  Acta Phys.\ Polon.\ B {\bf 39}, 1705 (2008)
  [arXiv:0803.4161 [hep-ph]].

\bibitem{Reuter:2016qbi} 
  J.~Reuter, B.~Chokoufe, A.~Hoang, W.~Kilian, M.~Stahlhofen, T.~Teubner and C.~Weiss,
  J.\ Phys.\ Conf.\ Ser.\  {\bf 762}, no. 1, 012059 (2016)
  [arXiv:1602.06270 [hep-ph]].

\bibitem{Frixione:1995ms} 
  S.~Frixione, Z.~Kunszt and A.~Signer,
  Nucl.\ Phys.\ B {\bf 467}, 399 (1996)
  [hep-ph/9512328].

\bibitem{Openloops} 
  F.~Cascioli, P.~Maierh\"ofer and S.~Pozzorini,
  Phys.\ Rev.\ Lett.\  {\bf 108}, 111601 (2012)
  [arXiv:1111.5206 [hep-ph]];
  F.~Buccioni, S.~Pozzorini and M.~Zoller,
  Eur.\ Phys.\ J.\ C {\bf 78}, no. 1, 70 (2018)
  [arXiv:1710.11452 [hep-ph]].

\bibitem{MPI}
  S.~Bra{\ss}, W.~Kilian, J.~Reuter,
  DESY 18-198, SI-HEP-2018-32, {\em in preparation}
  
\bibitem{Ohl:1998jn} 
  T.~Ohl,
  Comput.\ Phys.\ Commun.\  {\bf 120}, 13 (1999)
  [hep-ph/9806432].

\bibitem{Beneke:2015kwa} 
  M.~Beneke, Y.~Kiyo, P.~Marquard, A.~Penin, J.~Piclum and M.~Steinhauser,
  Phys.\ Rev.\ Lett.\  {\bf 115}, no. 19, 192001 (2015)
  [arXiv:1506.06864 [hep-ph]].

\bibitem{Hoang:2001mm} 
  A.~H.~Hoang, A.~V.~Manohar, I.~W.~Stewart and T.~Teubner,
  Phys.\ Rev.\ D {\bf 65}, 014014 (2002)
  [hep-ph/0107144].

\bibitem{Pineda:2006ri} 
  A.~Pineda and A.~Signer,
  Nucl.\ Phys.\ B {\bf 762}, 67 (2007)
  [hep-ph/0607239].
  
\bibitem{Hoang:2013uda} 
  A.~H.~Hoang and M.~Stahlhofen,
  JHEP {\bf 1405}, 121 (2014)
  [arXiv:1309.6323 [hep-ph]].

\bibitem{Jezabek:1992np} 
  M.~Jezabek, J.~H.~Kuhn and T.~Teubner,
  Z.\ Phys.\ C {\bf 56}, 653 (1992).

\bibitem{Harlander:1994ac} 
  R.~Harlander, M.~Jezabek, J.~H.~Kuhn and T.~Teubner,
  Phys.\ Lett.\ B {\bf 346}, 137 (1995)
  [hep-ph/9411395].

\bibitem{Hoang:1999zc} 
  A.~H.~Hoang and T.~Teubner,
  Phys.\ Rev.\ D {\bf 60}, 114027 (1999)
  [hep-ph/9904468].

\bibitem{Hoang:2010gu} 
  A.~H.~Hoang, C.~J.~Reisser and P.~Ruiz-Femenia,
  Phys.\ Rev.\ D {\bf 82}, 014005 (2010)
  [arXiv:1002.3223 [hep-ph]].
  
\end{thebibliography}
\end{document}